\documentclass[prl,twocolumn,showpacs,amsmath,amssymb,superscriptaddress]{revtex4-1}

\usepackage{amsmath,amssymb,amsfonts,bm,color,graphicx,tabularx,physics}

\usepackage[unicode=true,colorlinks=true]{hyperref}

\usepackage[etex=true,export]{adjustbox}

\hypersetup{linkcolor=blue,citecolor=blue,urlcolor=blue}
\usepackage{diagbox}

\newcommand{\s}{{\sigma}}

\newcommand{\beq}{\begin{equation}}
\newcommand{\eeq}{\end{equation}}
\newcommand{\beql}{\begin{equation*}}
\newcommand{\eeql}{\end{equation*}}
\newcommand{\beqn}{\begin{eqnarray}}
\newcommand{\eeqn}{\end{eqnarray}}

\newcommand{\sgn}{{\rm sgn}}

\renewcommand{\vec}[1]{\mbox{\boldmath$#1$}}

\begin{document}

\title{Braiding higher-order Majorana corner states through their spin degree of freedom}
	\author{Xiao-Hong Pan}
	\author{Xun-Jiang Luo}
	\affiliation{School of Physics and Institute for Quantum Science and Engineering, Huazhong University of Science and Technology, Wuhan, Hubei 430074, China}
	\affiliation{Wuhan National High Magnetic Field Center and Hubei Key Laboratory of Gravitation and Quantum Physics, Wuhan, Hubei 430074, China}
	\author{Jin-Hua Gao}
	\email{jinhua@hust.edu.cn}
	\affiliation{School of Physics and Institute for Quantum Science and Engineering, Huazhong University of Science and Technology, Wuhan, Hubei 430074, China}
	\affiliation{Wuhan National High Magnetic Field Center and Hubei Key Laboratory of Gravitation and Quantum Physics, Wuhan, Hubei 430074, China}
	\author{Xin Liu}
	\email{phyliuxin@hust.edu.cn}
	\affiliation{School of Physics and Institute for Quantum Science and Engineering, Huazhong University of Science and Technology, Wuhan, Hubei 430074, China}
	\affiliation{Wuhan National High Magnetic Field Center and Hubei Key Laboratory of Gravitation and Quantum Physics, Wuhan, Hubei 430074, China}
\date{\today}
\begin{abstract}
In this work, we study the spin texture of a class of higher-order topological superconductors (HOTSC) and show how it can be used to detect and braid Majorana corner modes (MCMs). This class of HOTSC is composed of two-dimensional topological insulators with s-wave superconductivity and in-plane magnetic fields, which offers advantages in experimental implementation. The spin polarization of the MCMs in this class is perpendicular with the applied magnetic field direction and is opposite on intrinsic orbitals, resulting in an overall ferrimagnetic spin texture. As a result, we find that the spin-selective Andreev reflection can be observed in a transverse instead of parallel direction to the applied magnetic field. Meanwhile, this spin texture leads to the gate-tunable $4\pi$ periodic $\phi_0$ Josephson current that performs qualitatively different behavior from the topologically trivial $\phi_0$-junction under rotating the in-plane magnetic field. Meanwhile, the existence of the MCMs in this class does not depend on the in-plane magnetic field direction. This gives rise to great advantage in constructing all electronically controlled Majorana network for braiding, which is confirmed through our numerical simulation. We thus provide a comprehensive scheme for probing non-Abelian statistics in this class of HOTSCs.

\end{abstract}
\maketitle
The braiding of Majorana zero modes (MZMs) is crucial for revealing their non-Abelian braiding statistics and implementing topological quantum computing (TQC)\cite{Ivanov2001,Kitaev2003,Nayak2008,Alicea2012,Beenakker2013,Sarma2015,Elliott2015}. In particular, it is now widely accepted that the zero-bias conductance peak (ZBCP) alone is insufficient to confirm the existence of MZM\cite{Pientka2012,Liu2012,Cole2016,Liu2017,DasSarma2021,Pan2021a,Pan2021}. Only the observation of non-Abelian braiding statistics is the most conclusive proof of their existence. Therefore, an experimentally feasible braiding and reading scheme is necessary for any promising Majorana platform.  Meanwhile, the MZM wave function is fully spin-polarized \cite{Sticlet2012,He2014,Sun2016,Serina2018,Glodzik2020}, resulting in universal spin-triplet superconducting correlation \cite{Asano2013,Ebisu2015,Liu2015,Zhang2017,Chen2019}. Accordingly, these spin properties can cause spin-dependent transport phenomena such as spin-selective Andreev reflection and topological $\phi_0$-junction \cite{Lutchyn2010,Badiane2011,He2014,Sun2016,Liu2016,Guigou2016,Prada2017,Aligia2020}, which can serve as the evidence of MZMs beyond the ZBCP. Remarkably, the latter comes from the spin-dependent Josephson coupling, which is not only useful in measuring the fermion-parity of two MZMs fusion but also essential to implement the MZMs braiding \cite{Alicea2011,Heck2012,Beenakker2013,Liu2014,Liu2016,Stern2019}. Although the spin properties of Majorana zero-energy modes can provide a significant route for their detection and braiding, their studies have primarily focused on first-order TSCs\cite{Vijay2016,Stern2019}. Recent studies on higher-order topological superconductors (HOTSCs) have opened up a new perspective: directly implementing MZMs on corners of two-dimensional (2D) or three-dimensional (3D) systems\cite{Langbehn2017,Yan2018,Wang2018a,Wang2018,Hsu2018,Liu2018,Volpez2019,Zhang2019,Zhu2019,Ezawa2019,Ghorashi2019,Franca2019,Pan2019,Zhang2019a,Yan2019a,Yan2019,Tiwari2020,Wu2020,Ghorashi2020,Zhang2020,Zhang2020a,Li2021a,Ghosh2021a,Ghosh2021,Li2021,Chen2021,Niu2021,Luo2021}. This not only provides a new topological phase but also has advantages for MZM experimental realization. Although many works have provided rich proposals for the realization of HOTSC, the braiding scheme of MZMs in HOTSC has not been well investigated.

\begin{figure}
\centering
\includegraphics[width=1\columnwidth]{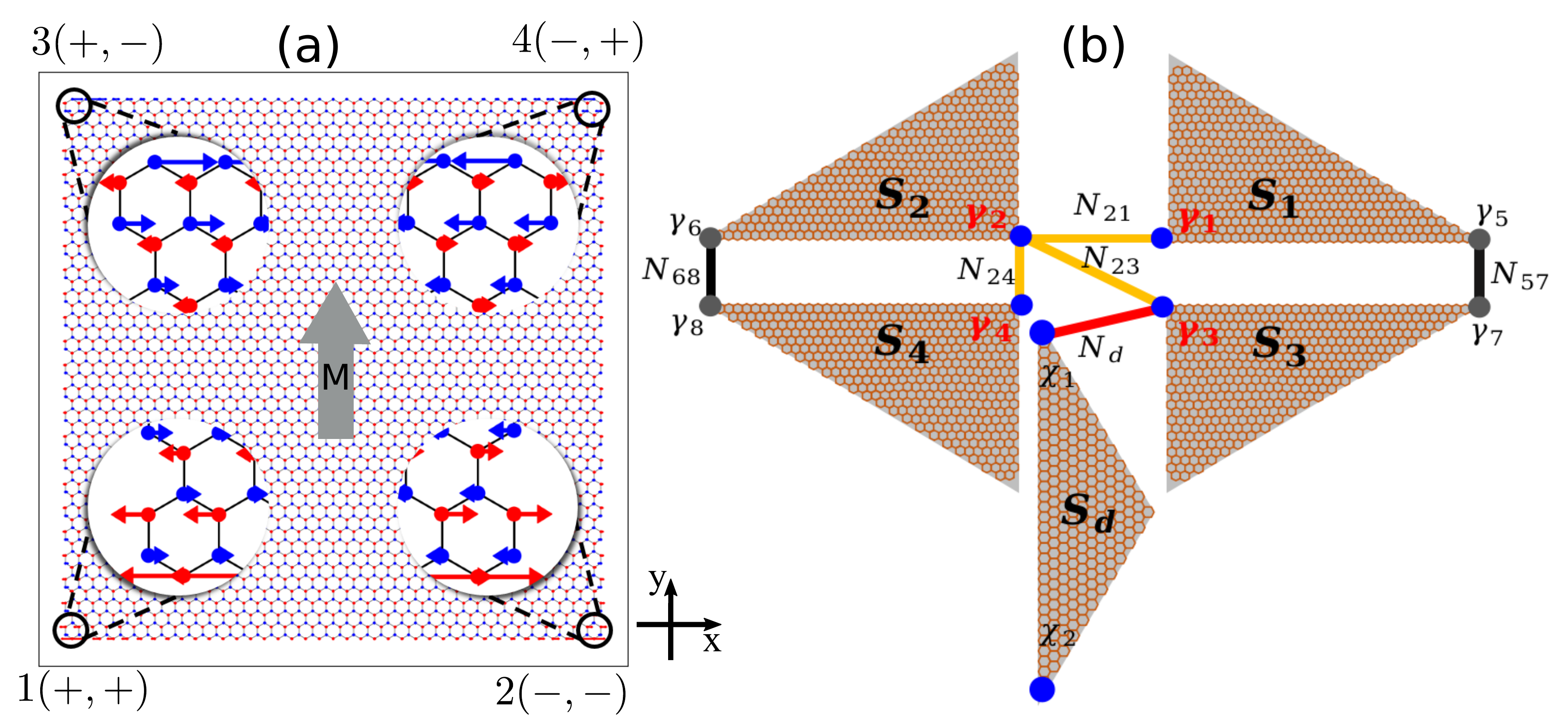}
\caption{(a) By fixing magnetic field along +y, the spin polarization distribution of MCMs in bismuthene when the system preserve chiral symmetry at parameters according to Ref.\cite{Li2018}. (b)   are HOTSCs. The MCMs $\gamma_{1,2,3,4},\chi_{1,2}$ sit on the HOTSC $S_{1,2,3,4,d}$. $N_{21}$, $N_{23}$, $N_{24}$ are the normal wires to control the MCMs coupling. The MCMs $\gamma_{5,6,7,8}$ remain coupled through the wires $N_{57}$ and $N_{68}$ in the whole process. $N_d$ is the normal wire connecting $\gamma_3$ and $\chi_{1}$ which are used to detect the braiding result.}
\label{Fig_1}
\end{figure} 

In this work, we study the spin texture of Majorana corner modes (MCM) in a class of HOTSC and show how it bring advantages in detection as well as implementation of braiding MCMs. The HOTSC is made up of a two-dimensional topological insulator \cite{Zhang2014,Li2018,Reis2021}, an s-wave superconductor, and an in-plane magnetic field, all of which are experimentally feasible. In the case that the system has well-defined chiral symmetry, we found that this class of HOTSC supports the MCMs whose spin direction is perpendicular to the magnetic field direction (Fig.~\ref{Fig_1}(a)). As a result, the spin-selective Andreev reflection can be observed in a transverse direction to the applied magnetic field. Remarkably, this spin texture leads to the anomalous $4\pi$ periodic $\phi_0$ Josephson current that performs qualitatively different from the topologically trivial $\phi_0$-junction. This is not only useful for detecting MCMs but also required for the implementation of all electrically controllable Majorana braiding. Meanwhile, the existence of MCMs in the class of HOTSC we study is independent on the in-plane magnetic field direction\cite{Pan2019}. This not only provides additional parameter space to distinguish the MCMs from trivial states but also bring significant advantages in the construction of the Majorana network (Fig.~\ref{Fig_1}(b)). Based on this network, we provide a comprehensive scheme for braiding MCMs and probing their non-Abelian statistics in HOTSCs through topological $\phi_0$-junction.

{\it Spin texture of MCMs} - The class of HOTSC we study can be generally described by a minimal Hamiltonian \cite{Pan2019,Wu2020,Zhang2020,Zhang2020a}
\beqn\label{Bulk-H}
H(k)&=&d_{1}(k)\tau_{z}s_{0}\sigma_{x}+d_{2}(k)\tau_{z}s_{0}\sigma_{y}+d_{3}(k)\tau_{0}s_{z}\sigma_{z}\nonumber \\
&&+M(\cos\theta \tau_z s_x + \sin\theta \tau_{0}s_{y})\sigma_0+\Delta\tau_ys_{y}\sigma_{0},
\eeqn
where the first three terms give the Hamiltonian of 2D TI with $d_{i=1,2,3}(k)$ depending on the specific models, such as Kane-Mele \cite{Kane2005a,Kane2005} and Bernevig-Hughes-Zhang models \cite{Andrei2021,Qi2011}, Pauli matrices $\tau$, $s$ and $\sigma$ act on the Nambu, spin and orbital \cite{supp} space respectively, $M$ and $\Delta_{\rm sc}$ are magnitude of Zeeman splitting and s-wave paring, respectively, and $\theta$ is the polar angle in the x-y plane indicating the direction of magnetic field. It has been shown in several works that the existence of MCMs does not depend on the in-plane magnetic field direction\cite{Pan2019}. For simplicity and without loss of generality, we first take the magnetic field along $y$ direction. The Hamiltonian of Eq.~\eqref{Bulk-H} possess two chiral symmetry\cite{supp}, and the operator in this case take the form
\beqn
 \mathcal{C}_{1}=\tau_{0}s_{x}\sigma_{z}\ \ \text{and}  \ \  \mathcal{C}_{2}=\tau_{y}s_{x}\sigma_{0}.
\eeqn

The two chiral symmetry operators commute with each other and are block diagonal in orbital space. Therefore we can classify their eight common eigen-basis $\Phi_{(\epsilon_1,\epsilon_2)}^{\lambda}$ into two categories according to $\lambda$ which is the eigenvalues of $\sigma_{z}$ in Tab.~\ref{tab-1}. 
 \begin{table}[htbp]
 \centering
  \caption{Eigen-basis of chiral operators. \label{tab-1}}
 \begin{tabular}{|c|c|c|}
 \hline
 \diagbox{$(\epsilon_1,\epsilon_2)$}{$\lambda$} & $+$ & $-$ \\
 \hline
 $(+,+)$& $ |\tau_y=+1\rangle\otimes|\rightarrow\rangle $ & $ |\tau_y=-1\rangle\otimes |\leftarrow\rangle$ \\
 \hline
 $(+,-)$ & $|\tau_y=-1\rangle\otimes|\rightarrow\rangle $ & $|\tau_y=+1\rangle\otimes |\leftarrow\rangle$ \\
 \hline
 $(-,-) $  & $ |\tau_y=+1\rangle\otimes|\leftarrow\rangle $ & $ |\tau_y=-1\rangle\otimes |\rightarrow\rangle$ \\
 \hline
 $ (-,+) $ & $|\tau_y=-1\rangle\otimes|\leftarrow\rangle $ & $|\tau_y=+1\rangle\otimes |\rightarrow\rangle$ \\
 \hline  
 \end{tabular}
\end{table}

\noindent Here, $\epsilon_1$ and $\epsilon_2$ are the eigenvalues of $\mathcal{C}_1$ and $\mathcal{C}_2$ respectively and $\rightarrow(\leftarrow)$ indicates the spin along positive (negative) $x$ direction.
Due to the chiral symmetry, the MCMs must be the eigenstates of the chiral operators. Consequently, each MCM can be labeled by the eigenvalues $(\epsilon_{1},\epsilon_{2})$. We note that the system also has effective mirror symmetry $\hat{M}_{x} H(k_x,k_y)\hat{M}_{x}^{-1} = H(-k_x,k_y)$ and $\hat{M}_{y} H(k_x,k_y)\hat{M}_{y}^{-1} = H(k_x,-k_y)$ with $\hat{M}_{x}=i\tau_{0}\sigma_{0}s_{y}$ and $\hat{M}_{y}=-i\tau_{0}\sigma_{x}s_{y}$ respectively. The product of these two effective mirror operators, $\hat{I}=\hat{M}_{x}\hat{M}_{y}=\tau_0\sigma_x\s_0$, gives the inversion operator $\hat{I} H(k_x,k_y) \hat{I}^{-1} = H(-k_x,-k_y)$. Without loss of generality, we take the MCM at corner 1 (Fig.~\ref{Fig_1}(a)) to has eigenvalues of the chiral operators $(+,+)$ as
\beqn
\Psi_{1}(x,y)=\alpha(x,y)\Phi^{+}_{(+,+)}+\beta(x,y)\Phi^{-}_{(+,+)},
\eeqn
where $\alpha(x,y)$ and $\beta(x,y)$ provide the spatial distribution of the wave function.
Then, the MCMs localized at other corners can be obtained as
\beqn\label{mirror}
(\Psi_{2}, \Psi_{3}, \Psi_{4})=(\hat{M}_{x}, \hat{M}_{y},\hat{I})\Psi_{1}.
\eeqn
Note that the spin polarization in the eight-basis is perpendicular to the magnetic field. This is very different from the spin polarization of MZMs in the semiconductor nanowire and vortex core, which is parallel with the magnetic field direction. To study the physical property of this spin texture, thereafter, we take the 2DTI to be bismuthene which has been experimentally observed with topological non-trivial band gap of 0.5eV\cite{Zhang2014,Li2018,Reis2021}. The spin polarization of each MCMs is numerically calculated and plot in Fig.~\ref{Fig_1}(a), in which the four MCMs $\gamma_{i=1,2,3,4}$ is consistent with our symmetry analysis. Here the $\lambda=+1$ and $\lambda=-1$ refer to the A (blue) and B (red) sublattice of honeycomb lattice (Fig.~\ref{Fig_1}(a)).
The above symmetry analysis can be applied to arbitrary in-plane magnetic field direction \cite{supp}.
\begin{figure}
\centering
\includegraphics[width=1\columnwidth]{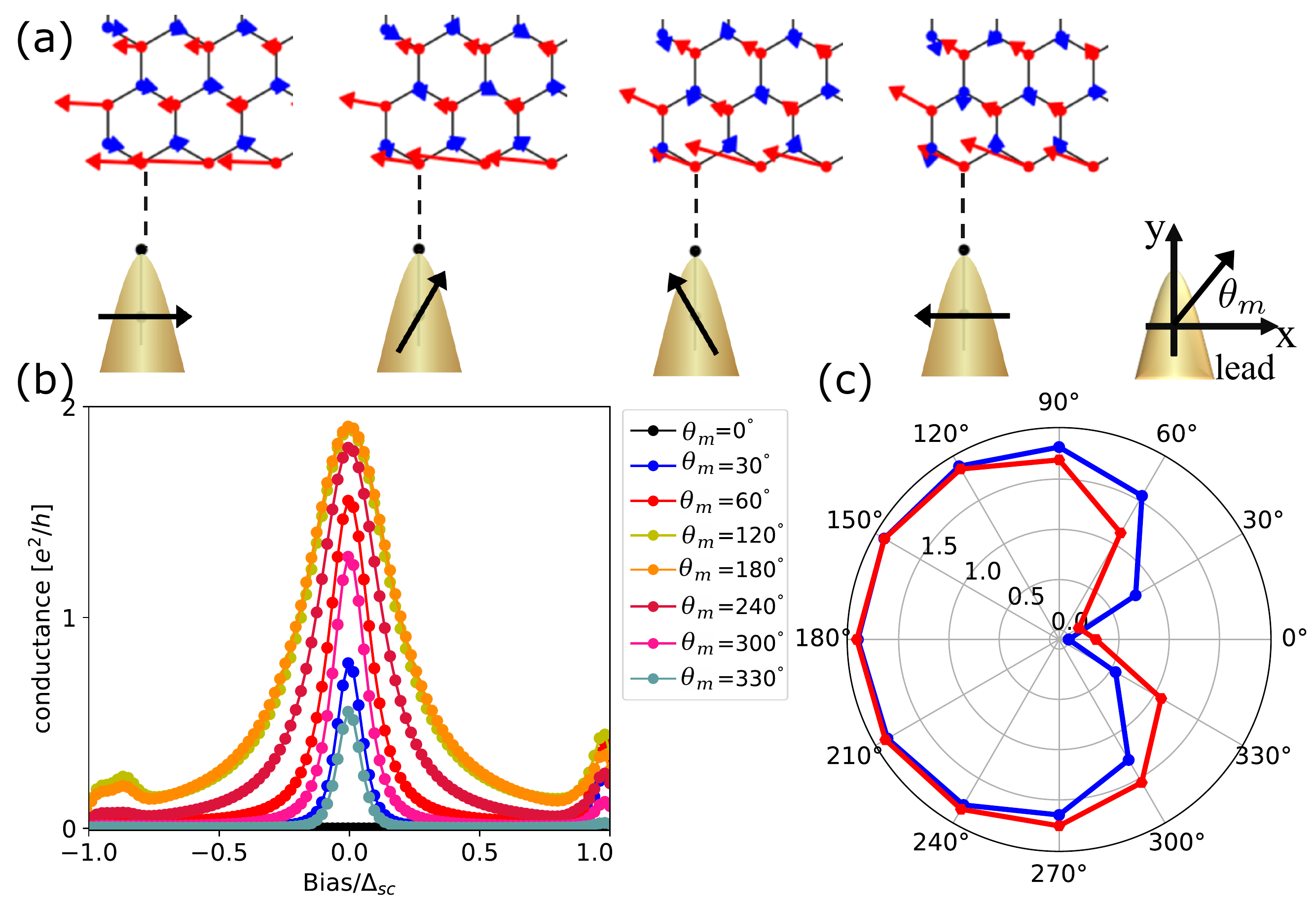}
\caption{The fixed magnetic field is along the +y direction. (a.)From left to right are the spin textures of the MCM at $\gamma_{\rm I}$ for lead polarizations along 0$^{\circ}$, 60$^{\circ}$, 120$^{\circ}$ and 180$^{\circ}$,while the system possess chiral symmetry. (b) is the conductance of MCMs at $\gamma_{\rm I}$ at $k_{B}T=0.03\Delta_{sc}$.(c)The blue and red curves represent the relationship between the polarization direction and the height of zero bias conductance, respectively. In order to reduce the computational effort, we mainly perform numerical calculations in low energy space of bismuthene model.}
\label{Fig_2}
\end{figure} 

\begin{figure*}[htbp]
\centering
\includegraphics[width=17.5cm]{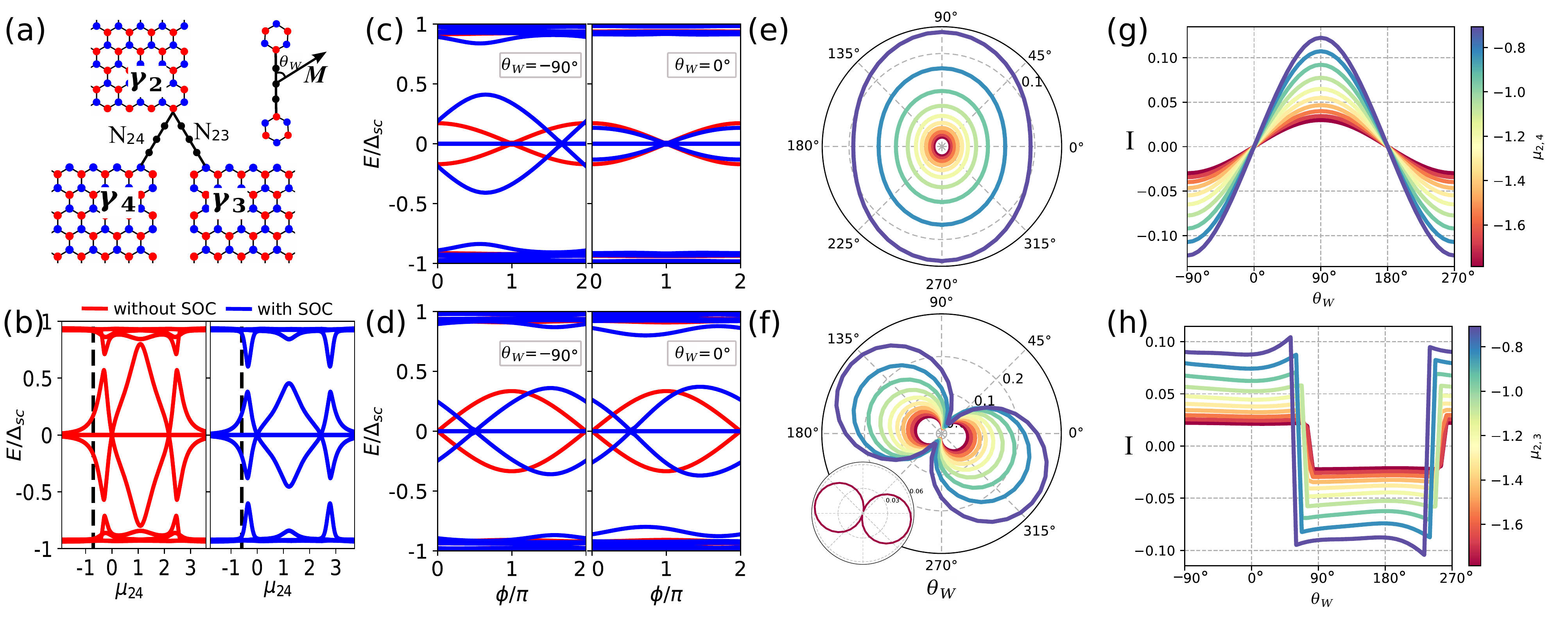}
\caption{(a)The configuration of the junction JJ$_{a}$ and JJ$_{b}$, and $\theta_W$ is the angle between in-plane magnetic field $\vec{M}$ and normal wire. (b) By fixing $\theta_W=-\pi/2$ and $\phi$=0, the red and blue curves are the energies change with chemical potential $\mu_{24}$ of the wires $N_{24}$ without and with SOC JJ$_{a}$ respectively. (c) Fixing chemical potential indicating by the black dot line in (b), the blue and red curves in the left and right figure, corresponding to $\theta_W=-\pi/2$ and 0, are the Andreev spectrum of JJ$_{a}$ in the case of normal wires without and with SOC respectively. (d) The left and the right figure are the Andreev spectrum of JJ$_{b}$ at $\theta_W=-\pi/2$ and 0, and the blue and red curves are representing the case of the wires without and with SOC. At $\phi$=0, (e)(f) are the dependence between Majorana couple and $\theta_W$ in JJ$_{a}$ and JJ$_{b}$, (g)(h) are the dependence between between current and $\theta_W$ in JJ$_{a}$ and JJ$_{b}$.}
\label{Fig_3}
\end{figure*}

{\it Transverse spin-selective tunneling} - 
The spin texture of the MCMs can be observed through observing the spin-selective Andreev reflection when the polarization of ferromagnetic lead is perpendicular with the magnetic field direction. We still first set magnetic field $\vec{M}$ to be along $y$ direction while the spin orientation of the MCM, $\gamma_{\rm I}$, at B sublattice is along -$x$ direction (Fig.~\ref{Fig_1}(a)). By attaching a ferromagnetic lead to the system (Fig.~\ref{Fig_2}(a)) and varying the polarization of the lead, we calculate the finite-temperature conductance as\cite{supp}
\beqn
G_{T}(V)=-\frac{2e^{2}}{\hbar}\int dE R(E)\frac{df(E-V)}{dE}
\eeqn
where $V$ is the bias voltage, $f$ is Fermi function and $R(E)$ is the probability of Andreev reflection obtaining from the transport package Kwant\cite{Groth2014}. The conductance at finite temperature reaches a minimum when the lead polarization is along $x$ direction, anti-parallel to the spin orientation of the MCMs without the lead. In this case, the lead has little effect on the orientation of MCMs (Fig~\ref{Fig_2}(a)). When the lead polarization deviates from this direction, the spin orientation of the MCM is deflected (Fig~\ref{Fig_2}(a)) while the conductance rises rapidly to the saturation value (Fig.~\ref{Fig_2}(b)). It can be see more obviously in our plot of the lead polarization dependent ZBCP in the polar coordinate (blue curves in Fig.~\ref{Fig_2}(c)). This is because the coupling between the spin-polarized lead and MCM break the local chiral symmetry and thus deflects the spin orientation of MCM (Fig.~\ref{Fig_2}(a)), which further enhance the coupling between the MCM and the lead. When the system itself breaks the chiral symmetry, the spin texture is not strictly but still roughly perpendicular to the direction of the magnetic field\cite{supp}. The lead polarization, corresponding to the minimal ZBCP, deviates from but is still close to $x$ direction (the red curves in Fig.~\ref{Fig_2}(c)). The high ZBCP shows similar behavior with that when the system respects chiral symmetry. Therefore, spin-selective Andreev reflection responds the applied magnetic field in the transverse direction, which is unique evidence for the MCMs in the class of HOTSC we are considering.

{\it Majorana coupling and topological $\phi_0$-junction} - 
Now, we consider two MCMs coupling through a normal lead (Fig.~\ref{Fig_3}(a)) and associated Josephson effect. Here, the two MCMs come from different HOTSCs, which form topological Josephson junctions. Without loss of generality, we fix one MCM to be $\gamma_2$, located at the right-bottom corner (Fig.~\ref{Fig_1}(a)). When considering the normal wire without SOC and taking the magnetic field perpendicular to the normal wire, we plot the Andreev levels of the Josephson junction connecting $\gamma_{2}$ with $\gamma_{4}$ and $\gamma_{3}$ respectively (red curves in Fig.~\ref{Fig_3}(c) and \ref{Fig_3}(d)). We find that the coupling between the two MCMs related by mirror operator, say $\gamma_2-\gamma_4$, and inversion operator, say $\gamma_2 - \gamma_3$, gives topological 0-junction and $\pi$-junction with $4\pi$ periodicity respectively. For convenience, we call the Josephson junction for the former and the latter case as JJ$_{\rm a}$ and JJ$_{\rm{b}}$. It has been shown that Majorana coupling is essential to perform fermion-parity measurements \cite{Vijay2016a,Karzig2017,Litinski2017} and MZMs braiding \cite{Sau2011,Beenakker2013,Pekker2013,Heck2015,Aasen2016,Stenger2019}. Meanwhile, fulfilling these tasks at $\phi=0$ can suppress the disturbing signals contributed from the possible topologically trivial channels. We then study the Majorana coupling magnitude at $\phi=0$. For JJ$_{a}$, we plot the eigenenergies at $\phi=0$ as a function of the chemical potential $\mu_{24}$ in the normal wire (in Fig.~\ref{Fig_3}(b)). The Majorana coupling, indicated by the lowest positive eigenenergy, increases exponentially in the low chemical region and then oscillates with further increasing $\mu_{24}$. Note that in the oscillation region, the coupling vanishes when $\mu_{24}$ across the eigenenergy of the isolated normal wire \cite{Liu2016}. Along with the requirement to turn on and off the Majorana coupling exponentially accurate, we focus on the low $\mu_{24}$ region. For JJ$_{\rm b}$, the Majorana coupling vanishes at $\phi=0$(red curves in Fig.~\ref{Fig_3}(d)), which is protected by the inversion symmetry \cite{supp}. To open the Majorana coupling at $\phi=0$ in the JJ$_b$, we add the Rashba SOC in the normal wire, which breaks the inversion symmetry and can be tuned by the electrical gate. The SOC shifts the Andreev crossing away from $\phi=0$ and therefore opens the Majorana coupling in JJ$_b$ (blue curves in Fig.~\ref{Fig_3}(d)). Meanwhile, for JJ$_{\rm a}$, the SOC also shifts the Andreev level crossing away from $\phi=\pi$ (blue curves in Fig.~\ref{Fig_3}(c)). Obviously, this shift results in topological $\phi_0$-junction with finite Josephson current at $\phi=0$. Here, as the existence of the MCMs does not depend on the in-plane magnetic field direction, we can study the Majorana coupling and associated Josephson current as a function of the magnetic field direction at $\phi=0$, which remarkably show very different behavior in JJ$_{\rm a}$ and JJ$_{\rm b}$. For JJ$_{\rm a}$, the Majorana coupling only slightly depends on magnetic field direction (Fig.~\ref{Fig_3}(e)). Meanwhile, the associated Josephson current amplitude oscillates with varying the angle between $\vec{M}$ and the normal wire. Particularly the current vanishes under all $\mu_{24}$ values for $\vec{M}$ parallel to the normal wire because the system restores the effective mirror symmetries in this case (Fig.~\ref{Fig_3}(g)). For JJ$_{\rm b}$, the Josephson coupling energy shows a strong $\vec{M}$ direction dependence which also depends on $\mu_{23}$. For very low $\mu_{23}$, the junction has the maximal Majorana coupling when the magnetic field is close to parallel with the normal wire (Fig.~\ref{Fig_3}(f)). With increasing $\mu_{23}$, the maximal coupling direction starts to gradually shift to the direction with an angle of 45° to the normal wire. Meanwhile, the associated Josephson current remains insensitive to the magnetic field direction in a large angle range and switches direction when the magnetic field is in near parallel with the normal wire. These numerical results can be understood semi-quantitatively when we project the Josephson junction Hamiltonian in MCMs Hilbert space, which gives the Majorana coupling Hamiltonian at $\phi=0$ 
\beqn
\tilde{H}_{23}&=& i\gamma_2 \gamma_3 \tilde{t}\sin\alpha_c \cos\frac{\beta}{2}, \nonumber \\
\tilde{H}_{24} &=& i\gamma_{2}\gamma_{4}\tilde{t} \cos(\theta_W+\alpha_c)\sin\frac{\beta}{2}
\eeqn
and the associative Josephson current
\beqn
I_{23} &=& \frac{e\Delta}{\hbar}\sin\theta_W\sin\frac{\beta}{2}, \nonumber \\
I_{24} &=& \frac{e\Delta}{\hbar}\sgn(\cos(\theta_W+\alpha_c)) \cos\frac{\beta}{2}.
\eeqn
where $\beta=\delta k L$, and $L$ is the length of the normal wire. Here, we assume the 1D Rashba SOC for simplicity\cite{supp}. It is clear that $\tilde{H}_{23}$ does not explicitly depends on magnetic field direction while the associative Josephson current does vanish at $\theta_W=0,\pi$ and reaches its maximal magnitude at $\theta_W=\pm \pi/2$. On the other hand, the Majorana coupling $\tilde{H}_{24}$ does explicitly depends on $\theta_W$ which can explain the strong magnetic field direction dependence of the coupling magnitude. Meanwhile, the associative Josephson current remains a constant and only change sign when $\theta_W+\alpha_c$ across $0$ or $\pi$. Note that $|\alpha_c|$ is larger when the coupling is stronger. The current switch deviates from the $\theta_W=\pm \pi/2$ more when the Majorana coupling is larger. These features are consistent with the numerical plot in Fig.~\ref{Fig_3}(g)(h). Moreover, it is worth noting that the Majorana coupling induced Josephson current in JJ$_a$ and JJ$_b$ behave very differently with varying the magnetic field direction. Meanwhile, when the system is in the topological trivial regime, the combination of the magnetic field and SOC can in principle also induces $\phi_0$-junction \cite{supp}. However, we found the induced Josephson currents in both JJ$_a$ and JJ$_b$ are very small and behave qualitatively different from those in JJ$_b$ \cite{supp}. Therefore, the observation of the $\phi_0$-junction in JJ$_b$ can nevertheless provides a strong evidence for system entry into the topological region.
 
\begin{figure}
\centering
\includegraphics[width=1\columnwidth]{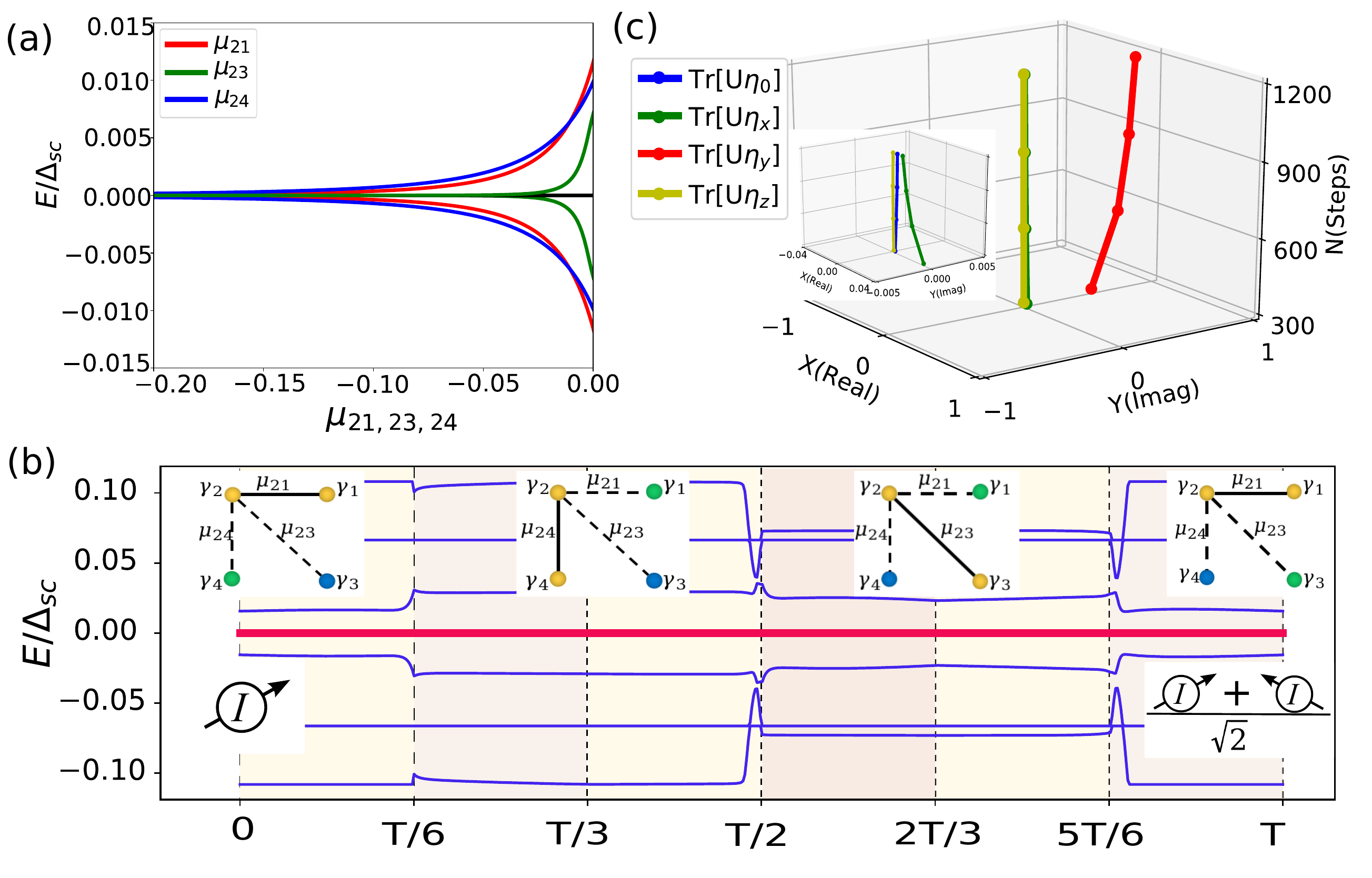}
\caption{(a)The relationship between chemical potential $\mu_{21}$, $\mu_{23}$, $\mu_{24}$ and the Majorana coupling magnitude respectively. (b) The energy spectrum in the whole procedure of exchanging $\gamma_{3}$ and $\gamma_{4}$. The insets plot characterizes the exchange protocol of $\gamma_{3}\leftrightarrow\gamma_{4}$. The ammeters indicate the Josephson current through $N_{d}$ for the initial and final states of the braiding process. (c)The relationship between the coefficients $(a_{0}, a_{x}, a_{y}, a_{z})$ and the total steps.}
\label{Fig_4}
\end{figure}

Now, we consider Majorana network shown in Fig.~\ref{Fig_1}(b) to braiding MCMs. The state is initialized by measuring the Josephson current direction through $N_{d}$. We then turn off the coupling between $\gamma_{3}$ and $\chi_1$ and start to exchange the MCMs $\gamma_3$ and $\gamma_4$ by exponentially turning on and off the Majorana couplings through tuning the chemical potential in the region shown in Fig.~\ref{Fig_4}(a). The whole process is divided into three intervals. In the first interval, We tune $\mu_{24}$ from 0 to $\mu_{24}^{\rm max}$ and $\mu_{21}$ from $\mu_{21}^{\rm max}$ to 0 so that the MCM at $\gamma_{4}$ position is moved to $\gamma_{1}$ position. In the second interval, we tune $\mu_{23}$ from 0 to $\mu_{23}^{\rm max}$ and $\mu_{24}$ from $\mu_{24}^{\rm max}$ to 0 so that the MCM at $\gamma_{3}$ position is moved to $\gamma_{4}$ position. At last, we change $\mu_{21}$ from 0 to $\mu_{21}^{\rm max}$ and $\mu_{23}$ from $\mu_{23}^{\rm max}$ to 0 so that the MCM at $\gamma_{1}$ position is moved to $\gamma_3$ position. During the braiding process, the energy of the two MZMs remains zero as indicated by the red line in Fig.~\ref{Fig_4}(b) and separated from excited states by a finite energy gap. 
The non-abelian berry phase accumulated in the braiding process can be revealed through the path-ordered Wilson loop calculation in the MCMs basis $(\Psi_{3}(t),\Psi_{4}(t))^{\rm T}$ \cite{Wilczek1984,Yu2011,Snizhko2019}
\beqn
U&=&\mathcal{P}e^{-\oint d\mu A(t)} = \prod_{n=0}^{N-1}\sum_{\alpha,\beta=3}^{4} \bra{\Psi_{\alpha}(t_{n})}\ket{\Psi_{\beta}(t_{n+1})}\nonumber \\
&=& a_{0} \eta_0 + a_x\eta_x + a_y\eta_y + a_z\eta_z,
\eeqn
where $t_{n+1}-t_{n}=\Delta t=T/N$, $N$ is the total steps for exchanging the MCMs in period $T$ and $\eta$ is Pauli matrix acting on the basis $(\Psi_{3}(0),\Psi_{4}(0))^{\rm T}$. Our numerical calculation shows that the braiding matrix $U$ has only $\eta_y$ component whose amplitude approaches $i$ with increasing the total steps $N$ (Fig.~\ref{Fig_4}(c)). This is consistent with the well know MZMs exchanging $\gamma_3 \rightarrow -\gamma_4$ and $\gamma_4 \rightarrow \gamma_3$ and renders the success of our all-electronically controlled braiding proposal. The experimental detection can be performed through measuring the Josephson current before and after the braiding. If the state is initialized by measuring the Josephson current direction through N$_d$, it will have 50\% probability to observe the current direction switch after the successful MCMs braiding \cite{supp}.

{\it Conclusion} - In conclusion, we provide a comprehensive scheme to detect and braiding MCMs in a class of HOTSC with the help of their unique spin properties. Our scheme is shown with the Kane-Mele model for realistic bismuthene but also valid for BHZ-model-base materials. The spin texture with the transverse polarization to the applied magnetic field leads to its spin-dependent transport qualitatively different from those in both the first-order TSC and topologically trivial superconducting states. This can provide the strong evidence that the system is in the HOTSC phase. We then propose an all electronically controlled Majorana network and numerically simulate the braiding process.


%

\end{document}